\documentclass[english,a4paper,prl,twocolumn,amsmath,amssymb,longbibliography,superscriptaddress,nofootinbib]{revtex4-2}
\usepackage{graphicx,color}
\usepackage{physics}
\usepackage{siunitx}
\usepackage[colorlinks,linkcolor=magenta,citecolor=magenta]{hyperref}

\newcommand{\mean}[1]{\langle #1 \rangle}

\newcommand{\IInt}[3]{\int_{#2}^{#3}\dd #1\;}
\renewcommand{\vec}[1]{\vb{#1}}
\newcommand{\pd}[2]{\frac{\partial #1}{\partial #2}}
\newcommand{\fd}[2]{\frac{\delta #1}{\delta #2}}

\newcommand{\al}{\alpha}
\newcommand{\gam}{\gamma}

\newcommand{\kap}{\kappa}

\newcommand{\vhi}{\varphi}
\newcommand{\sig}{\sigma}
\newcommand{\om}{\omega}

\newcommand{\Dr}{D_\text{r}}
\newcommand{\Dt}{D_\text{t}}

\newcommand{\kT}{k_\text{B}T}

\AtBeginDocument{%
    \newwrite\bibnotes
    \def\bibnotesext{Notes.bib}
    \immediate\openout\bibnotes=\jobname\bibnotesext
    \immediate\write\bibnotes{@CONTROL{REVTEX41Control}}
    \immediate\write\bibnotes{@CONTROL{%
    apsrev41Control,author="08",editor="1",pages="1",title="0",year="1"}}
     \if@filesw
     \immediate\write\@auxout{\string\citation{apsrev41Control}}%
    \fi
}%

\begin{document}
\title{Negative drag force on beating flagellar-shaped bodies in active fluids}

\author{Timo Knippenberg}
\affiliation{Fachbereich Physik, Universität Konstanz, 78464 Konstanz, Germany}
\author{Robin Bebon}
\affiliation{Institute for Theoretical Physics IV, University of Stuttgart, 70569 Stuttgart, Germany}
\author{Thomas Speck}
\email[Corresponding author: ]{thomas.speck@itp4.uni-stuttgart.de}
\affiliation{Institute for Theoretical Physics IV, University of Stuttgart, 70569 Stuttgart, Germany}
\author{Clemens Bechinger}
\email[Corresponding author: ]{clemens.bechinger@uni-konstanz.de}
\affiliation{Fachbereich Physik, Universität Konstanz, 78464 Konstanz, Germany}
\affiliation{Centre for the Advanced Study of Collective Behaviour, Universtätsstraße 10, 78464 Konstanz, Germany}

\begin{abstract}

We experimentally investigate the drag force exerted by a suspension of light-induced active particles (APs) on a translating and beating idealized flagellum-shaped object realized through negative phototactic interactions with the APs. We observe both positive and negative drag forces, depending on the beating frequency and translational velocity, driven by the dynamic redistribution of APs in response to the object's motion. These findings are supported by numerical simulations and an analytical model, extendable to a range of slender geometries. Our results illustrate the complex interplay between geometric body changes and the density distribution in active baths, which may also be relevant for microrobotic applications.
\end{abstract}

\maketitle


Active fluids such as bacterial suspensions, active colloidal particles (APs), or cell tissues, are intrinsically out of equilibrium and therefore exhibit fundamentally different dynamical properties and statistical behaviors compared to classical passive fluids~\cite{bechinger16}.
A particularly intriguing aspect of such systems arises in the presence of confinements or boundaries that have a strong impact on their spatial density distribution and collective organization. Geometric constraints typically lead to a curvature-dependent accumulation of bacteria and APs near walls and the emergence of surface-guided flows~\cite{nikola16,fily14,fily15}.

Such behavior of active fluids is a result of the self-propulsion of their constituents, which leads to forces acting onto walls and obstacles~\cite{smallenburg15}. When interacting with Brownian particles ~\cite{paul22,yamchi17,liu20,feng21}, such AP-generated forces lead to superdiffusive~\cite{wu00, leptos09, ortlieb19, kanazawa20, tripathi22, granek22} and even persistent motion of objects, both being absent in passive fluids~\cite{RN42, sokolov10, dileonardo10, yan15, speck20, speck21a}. Even more complex behaviors occur when rigid objects are externally driven through active fluids. When an object is subjected to a constant force or trapped in a moving harmonic trap, its effective friction is predicted to be reduced by more than an order of magnitude compared to a passive medium~\cite{burkholder20, knezevic21, solon22, jayaram23}. Such active thinning is due a distortion of the active fluid's microstructure, which in turn influences the reactive forces exerted from the fluid onto the object. Apart from a change of the fluctuation spectrum~\cite{RN33}, however, so far no experimental evidence for active thinning has been reported.

Here, we experimentally and theoretically investigate the force exerted by colloidal APs onto a periodically deforming and translating object mimicking the beating dynamics of a flagellum. Under certain conditions, we find that the drag force on the flagellum becomes negative, thus pointing along the flagellum's direction of motion. This is caused by the formation of spatio-temporal density distributions of the APs promoting the flagellar translation. Our experimental results are supported by numerical simulations but also confirmed by an analytical model that can be easily extended to estimate the mean forces exerted by an active fluid on slender immersed bodies, applicable in the limits of small and large transversal speeds.

In our experiments, we use an active fluid composed of light-propelled APs fabricated from optically transparent silica colloids with diameter $\sigma = \SI{7.89}{\micro\meter}$. The particles are coated on one hemisphere with a thin light-absorbing carbon layer. The APs are suspended in a mixture of water and propylene glycol n-propyl ether (PnP) at its critical composition, kept slightly below its critical demixing temperature of $T_\text{C}= \SI{32}{\celsius}$. When these particles are illuminated with laser light ($\lambda= \SI{532}{\nano\meter}$) the carbon caps are selectively heated above $T_\text{C}$, which leads to local demixing and eventually to self-propulsion \cite{RN4}. The propulsion velocity $v_0$ is adjustable by the incident light intensity. Due to gravity, the APs sediment towards the bottom plate, where they perform a two-dimensional (2D) active Brownian motion. The packing fraction of APs in our study has been adjusted to $\phi \approx 0.03$.

\begin{figure}[t]
    \centering
    \includegraphics{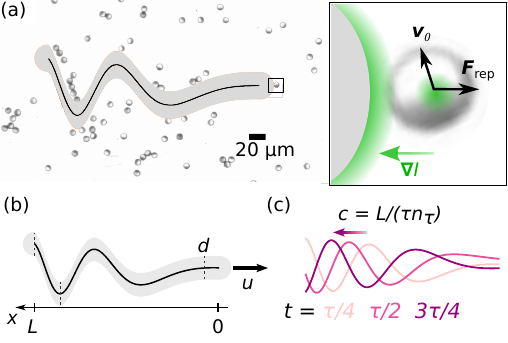}
    \caption{(a) Microscope image of APs expelled from a flagellum-shaped area (shaded in gray). The black solid line marks Eq.~\eqref{eq:flagellum_model}. Inset: Magnified view of an AP at the flagellum's edge. The corresponding light pattern giving rise to self-propulsion and repulsion is shown in green. (b) Contour and center of the virtual flagellum. The flagellum's length $L$, thickness $d$, and direction of transversal motion $u$ are indicated. (c) The undulation of the flagellum according to Eq.~\eqref{eq:flagellum_model} leads to periodic deformations, shown at different times.}
    \label{fig:model_sketch}
\end{figure}

To make the APs interact with the translating flagellar-shaped object, we exploit their negative phototactic behavior. When subjected to a light gradient $\vec{\nabla}I$, the APs experience a repulsive force $\vec{F}_\text{rep}~=~-f_0 \left( \vec{\nabla}I/|\vec{\nabla}I| \right)$, where the magnitude $f_0$ is related to the light intensity \cite{RN7}. In our experiments, $f_0 \approx 650 \; \kT/\sigma$, which is about three times the self-propulsion force of our APs (see Supplemental Material~\cite{sm} for details). Although our approach relies on the APs' negative phototaxis and the translating flagellum-shaped object is not a physical body, the repulsive forces experienced by the APs remain physical. This renders our experimental setup a suitable model system for investigating how boundary geometry influences interactions with active matter. To model the presence of the translating object, we create a dynamic light pattern which has the shape of a beating flagellum moving at a constant speed $u$ and from which the APs are repelled (gray area in Fig.~\ref{fig:model_sketch}(a)). We choose the direction of the $x$-axis to align with the opposite swimming direction so that $x=0$ corresponds to the head and $x=L$ to the tip of the flagellum's tail. See the supplemental material~\cite{sm} for more details.


The specific contour and beating dynamics of the flagellum-shaped object used in our study are motivated by an idealized eukaryotic flagellum model, which originally was proposed to describe motile bull spermatozoa~\cite{rikmenspoel65,machin58}. The time-dependent shape of the oscillating flagellum is given by the perpendicular displacement
\begin{align}
    \label{eq:flagellum_model}
    y(x,t) &= A_0 \left(1 + \frac{3x}{L}\right) \sin\left( \omega t + 2\pi n_\tau\frac{x^2}{L^2} \right)
\end{align}
from a straight line  illustrated in Fig.~\ref{fig:model_sketch}(b).
Its amplitude $A_0$ and length $L$ are chosen as $A_0 = \SI{10}{\micro\meter} \approx 1.25 \sigma $ and $L = \SI{275}{\micro\meter} \approx 35 \sigma$, yielding a similar $A_0/L$ ratio as in Ref.~\cite{rikmenspoel65}. Equation \eqref{eq:flagellum_model} describes the motion of a transversal wave that propagates along the flagellum with an average celerity $c = L/(\tau n_\tau)$ in backward ($+x$) direction. The period is $\tau = 2\pi/\om$ and the number of oscillations is $n_\tau = 1.75$. In Fig.~\ref{fig:model_sketch}(c), the shape of the propagating wave is illustrated for different phase angles. To account for the finite spatial extension in $y$-direction, we consider a thickness of $d = \SI{35}{\micro\meter} \approx 4.4 \sigma $. In addition to the beating pattern, the entire flagellum is moved at constant speed $u$ along the negative $x$-direction, and we consider the flagellum's comoving frame henceforth (see also Supplementary Movie 1). We note that our experimental approach neglects possible hydrodynamic interactions between APs and a physical body. Similar to \emph{E.~coli}, however, whose interaction with walls are dominated by short-range steric forces~\cite{drescher11}, AP--wall interactions in our system have been shown to be weak~\cite{volpe11}. We stress that the experimental setup fully accounts for hydrodynamic AP--AP interactions.


\begin{figure}[t]
    \centering
    \includegraphics{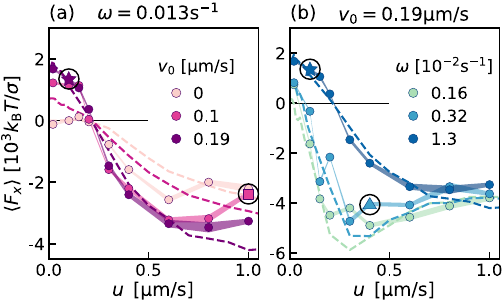}
    \caption{(a,b) The $x$-component of the average force that the APs exert on the flagellum $\langle F_x \rangle$ at an AP area packing fraction of $\phi \approx 0.03$ as a function of transversal speed $u$. The symbols are experimentally measured data, the shaded areas indicate the corresponding standard deviation of mean. The dashed lines represent simulation results for same parameters as in experiments. In (a), the flagellum's beating frequency is kept constant at $\omega = \SI{0.013}{\per \second}$ and the APs' speed $v_0$ is varied. In (b) $v_0 = \SI{0.19}{\micro\meter\per\second}$ and $\omega$ is varied.}
    \label{fig:force_x}
\end{figure}

To measure the mean force $\langle \vec F \rangle$ experienced by the flagellum-shaped body within a bath of APs, we take advantage of the fact that the entire system would be force-free. The repulsive force $\vec{F}_{\text{rep},i}$ experienced by a single AP is balanced by an opposite force on the flagellum 
\begin{equation}
    \langle \vec{F} \rangle = \left\langle - \sum_{i=1}^{N_t} 
    \vec{F}_{\text{rep},i}(t)  \right\rangle_t ,
    \label{eq:force:exp}
\end{equation}
which we can measure and average over the duration of an integer number of full oscillations. Here, $N_t$ is the number of APs that experience a repulsive force from the flagellum at time $t$, see End Matter and SM \cite{sm} for further details on the force measurement and the averaging.
Due to the symmetry of Eq.~\eqref{eq:flagellum_model}, the average force on the flagellum perpendicular to its direction of motion vanishes (exemplary measurements are shown in the SM \cite{sm}), and only the $x$-component $\langle{F_x}\rangle$ of Eq.~\eqref{eq:force:exp} is considered in the following. 

Figure~\ref{fig:force_x}(a,b) shows experimentally obtained forces $\langle F_x \rangle$ as a function of the flagellum’s transversal speed $u$ plotted for different propulsion speeds $v_0$ and beating-frequencies $\omega$. In case of a passive bath ($v_0=0$), the flagellum experiences almost no drag for small speeds $u$. This is caused by the periodic beating of the flagellum, which pushes APs out in the perpendicular direction and thus only few APs enter the envelope of the oscillating wave pattern. These exert a force on the flagellum while being transported to the rear end, which is compensated by the drag experienced at the front. For larger speeds, the drag becomes dominant and $\langle F_x \rangle<0$ opposes the translational motion of the flagellum, similar to Stokes-friction in a homogeneous liquid (Fig.~\ref{fig:force_x}(a)). For $v_0 \neq 0$ the force $\langle F_x \rangle >0$ at small speeds $u$ increases appreciably and the flagellum experiences a negative drag, i.e. its motion is facilitated by the interactions with the surrounding APs (Fig.~\ref{fig:force_x}(a)).
A rough estimate yields that this negative drag is more than four times larger than the Stokes friction force of a sphere with radius $A_0$ (see~\cite{sm}).

Further increasing the translational speed $u$ reduces the force until it changes sign at $u_\ast$, which corresponds to the speed that a free flagellum would move at. Remarkably, for fixed beating frequency, the value of $u_\ast$ is independent of $v_0$ and identical to the speed beyond which negative forces are observed in the passive case. For speeds $u>u_\ast$, the force $\langle F_x \rangle<0$ resembles the behavior of a passive environment ($v_0=0$). The same behavior for the force $\langle F_x \rangle$ is found for other values of $\omega$ with $u_\ast$ depending on $\omega$ but not on $v_0$ (Fig.~\ref{fig:force_x}(b)).

\begin{figure}
    \centering
    \includegraphics{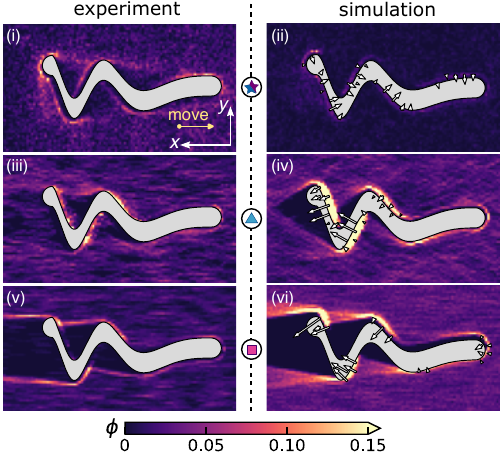}
    \caption{Experimentally (left) and numerically (right) determined AP density maps at fixed phase angle $\simeq 0.54 \pi$ of the flagellar beating. AP densities are accumulated over a time window corresponding to $0.05 \pi$. Simulation panels additionally show the direction and magnitude of $\vec{F}$ marked as white arrows. The respective parameters are indicated by the highlighted star (a,b), square (a), and triangle (b) symbols in Fig.~\ref{fig:force_x}.}
    \label{fig:heatmap}
\end{figure}

To elucidate the origin of these forces, the left column of Fig.~\ref{fig:heatmap} shows the experimentally measured density maps of APs in the vicinity of the flagellum. For $u<u_\ast$, APs accumulate preferentially at the back of the flagellum (i). This explains why at small $u$ the translation of the flagellum becomes facilitated in the presence of APs ($\langle F_x \rangle > 0$). In contrast, for $u > u_\ast$ dense regions of APs form primarily at the front of the traveling flagellum (iii). This impedes its translation and leads to $\langle F_x \rangle < 0$. For $u \gg u_\ast$ pronounced depletion regions form around the posterior part of the flagellum (v). This leads to a further accumulation of APs at the front similar to (iii), but with a slightly decreased accumulation area.

The data shown are corroborated by Brownian dynamics (BD) simulations, in which the APs are modeled as active Brownian particles moving in 2D (see~\cite{sm}). The corresponding BD results are shown as dashed lines in Fig.~\ref{fig:force_x}(a,b) and as the right column of Fig.~\ref{fig:heatmap}, respectively. Both, the drag forces $\langle F_x \rangle$ and the AP density distributions around the flagellum show good agreement with our experiments. We emphasize that the observed features qualitatively do not depend on details of the flagellum model. We tested this by performing simulations with two different shapes, yielding qualitatively similar results (see~\cite{sm}).

The comparison of $\langle F_x \rangle$ with the corresponding density maps shows that the sign and magnitude of the drag force on the flagellum is determined by the spatial distribution of the APs along its contour. Opposed to a static body, however, where the local density of surrounding APs is essentially determined by its local curvature, the AP density of a translating and beating flagellum is much more complex. This is due to the competition of the time scales characterizing the flagellum and the active bath. Based on these observations, we now develop a simple theoretical model that is able to capture the general observed phenomena.

\begin{figure*}
    \centering
    \includegraphics{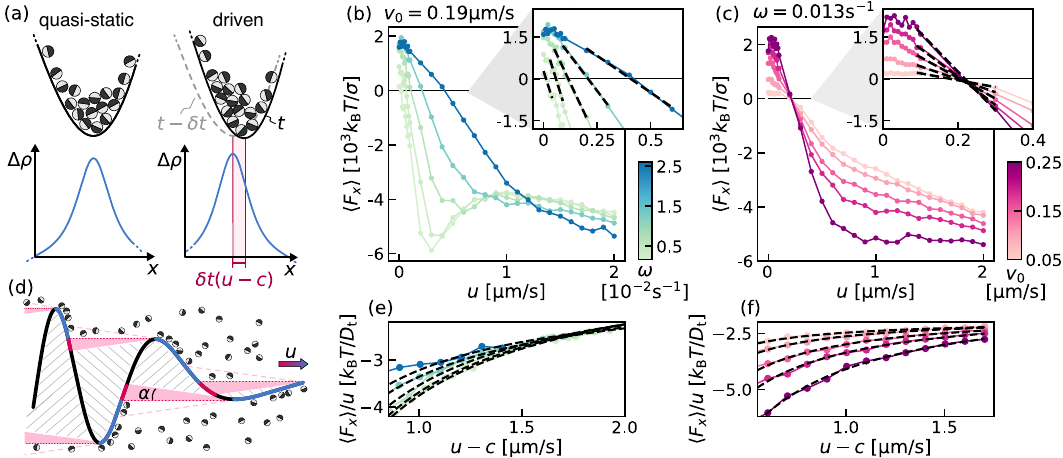}
    \caption{(a) Schematic of how the local particle distribution differs between the quasi-static and driven case. (b,c) The $x$-component of forces measured in simulations for: (b) different beating frequencies $\om$ at fixed $v_0 = \SI{0.19}{\micro\meter\per\second}$ and (c) different bath particle speeds $v_0$ at fixed $\om = \SI{0.013}{\per\second}$. Insets respectively show zooms of the zero crossing where dashed lines indicate fits of the theoretical prediction. (d) Schematic wake formation at intermediate to large $u$. Blue segments depict the accessible regions in the asymptotic $u \to \infty$ limit and red shaded regions indicate the increased accessibility for intermediate speeds with non-zero ratios $v_0/(u-c)$. (e,f) Scaled forces $\langle F_x \rangle/u$ for large $u$, fitted with the theoretical prediction from Eq.~\eqref{eq:fit} (dashed lines).}
    \label{fig:force_sim}
\end{figure*}

Our simplified analytical model is based on the fact that the force exerted on an immersed object is determined by the aggregation of APs along its body. We denote the densities of APs within the accumulation layer on the upper and lower sides of the body by $\rho_+(x,t)$ and $\rho_-(x,t)$ respectively. The local force density
\begin{equation}
    \vec{f}(x,t) = -f_0\vec{n}(\rho_+ - \rho_-) = -f_0\vec{n}\Delta\rho
    \label{eq:force}
\end{equation}
is then governed by the density difference $\Delta \rho \equiv \rho_+ - \rho_-$ across the body. The normal vector $\vec n = (-\sin\theta, \cos\theta)^\top$ is chosen such that forces aligning with the flagellum take positive values. The local angle $\theta(x,t)$ is defined via $\tan\theta(x,t)=y'(x,t)$, where the prime denotes the derivative with respect to $x$. The mean force exerted on the flagellum per period
\begin{equation}
    \mean{\vec{F}} = \frac{1}{\tau}\IInt{t}{0}{\tau}\IInt{x}{0}{L}\vec{f}(x,t)
    \label{eq:force_Integrated}
\end{equation}
is determined through integrating Eq.~\eqref{eq:force} along the full length $L$, followed by temporal averaging over one period $\tau = 2\pi/\om$.

To proceed, we require an explicit expression for $\vec{f}(x,t)$. In the limit of quasi-static weak driving, we draw inspiration from the study of active Brownian particles in presence of curved and deformable confinements~\cite{fily14, smallenburg15, fily15, duzgun18, wang19, nikola16, diaz24} and posit that the density difference $\Delta \rho \approx \zeta \kap$ is due to motility-induced wall aggregation and thus directly proportional to the signed curvature,
$ \kappa(x,t) = y''/(1+y'^2)^{3/2}$,
ignoring effects stemming from the finite size of the body. The coefficient $\zeta$ depends on details of the interactions between APs and flagellum and is sensitive to the choice of parameters $v_0$ and $\om$. A numerical validation of this assumption is provided in~\cite{sm}. The asymmetry inherent in the position-dependent amplitude and phase of $y(x,t)$ yields a negative force component $\mean{F_x}$ while $\mean{F_y}$ vanishes after integration, both in agreement with physical intuition and simulation results.

When moving the body with speed $u$, the experiments have revealed the crucial role of the competing time scales that characterize the flagellum's beating relative to the relaxation of the active bath. More specifically, the redistribution of APs in the accumulation layer is determined by the relative speed $u-c$ experienced by a fixed point on the flagellum's contour. For $u \approx c$ the curvature remains (almost) constant at each point in the lab reference frame, permitting APs to aggregate preferentially in regions with high curvature. For $u \neq c$, APs experience a more pronounced change in their local environment as the body's contour shifts with speed $u-c$. The redistribution of APs inside concave regions occurs on time scales $\delta t \ll D_\text{r}^{-1}$ much shorter than the one associated with diffusive reorientation, given by the rotational diffusion coefficient $D_\text{r}$. We account for the time $\delta t$ that the density difference lags behind through $\Delta\rho(x,t) \approx \zeta \kappa(x', t')$, where the curvature is now evaluated at an earlier time $t'=t-\delta t$ at corresponding position $x' = x+(u-c)\delta t$. An expansion for small lag times $\delta t$ yields
\begin{align}
    \Delta \rho \approx \zeta\kappa - \zeta\delta t \left[\partial_t \kappa + \left(c -u\right)  \partial_x \kappa  \right].
    \label{eq:curv_driven}
\end{align}
For a schematic comparison between the AP accumulation in the quasi-static and driven case, see Fig.~\ref{fig:force_sim}(a). To validate our theory, we compare it with extensive simulation data presented in Fig.~\ref{fig:force_sim}(b,c) that explores a wider range of parameters than accessible in the experiments. Substituting Eq.~\eqref{eq:curv_driven} into Eq.~\eqref{eq:force}, we numerically integrate and fit parameters $\zeta$ and $\delta t$ of the resulting average force (Eq.~\eqref{eq:force_Integrated}) to obtain the dashed lines depicted in the insets of Fig.~\ref{fig:force_sim}(b) and (c). 
Our minimal ansatz of delayed curvature-driven aggregation thus reproduces the primary characteristics for the drag force found in experiments and simulations: The swim speed $u_\ast$ depends linearly on the beating frequency (cf.~Fig.~S4(a) in~\cite{sm}) and is independent of bath activity. The latter, however, determines the force amplitude.

In the opposite limit of very large speeds $u\to \infty$, APs have essentially no time to diffuse into the concave regions and the force reduces to the linear, Stokes-like dependency $\mean{F_x} \simeq -\gam_\infty u$ with the asymptotic drag coefficient $\gam_\infty$. At intermediate translational speeds, V-shaped wakes form at each extremum of the periodic body shape, increasing the extent of the accumulation layer. The increased coverage can even lead to local minima in the force (see Fig.~\ref{fig:force_x}(b) and Fig.~\ref{fig:force_sim}(b)). Following a simple geometric construction, the opening angle of each wake is $2\al = 2\arctan(v_0/(u-c))$. Per time $\Delta t$ the APs move perpendicular at most a distance $\Delta y = \pm v_0 \Delta t$, while a given point on the body covers the distance $\Delta x = (u-c) \Delta t$ along with the body. The triangle spanned by $(\Delta x, \Delta y)^\top$ and $\Delta x \vec{e}_x$ opens with angle $\al$ (marked as red shaded regions in Fig.~\ref{fig:force_sim}(d)) and the cross section of the body increases proportional to $\sin(\al)$ (projection of dark red segments) when compared with the $u \to \infty$ limit (blue segments). Collecting terms, we predict that the average force behaves as
\begin{equation}
    \mean{F_x} \simeq -\gam_\infty u (1 + \beta \sin\al)
    \label{eq:fit}
\end{equation}
for intermediate to large object speeds, where $\beta$ is a free parameter. We find values for $\beta$ in the range of 5 to 20, which increase with $v_0$ and decrease with $\omega$. Plotting the scaled force $\mean{F_x}/u$ in Fig.~\ref{fig:force_sim}(e,f) shows excellent agreement between data and theory. Notably, while curves of equal bath activity collapse in the limit of large $u$ (Fig.~\ref{fig:force_sim}(e)), varying $v_0$ results in a wider spread (Fig.~\ref{fig:force_sim}(f)) since $\gam_\infty$ reduces with increased bath activity. In this shear-thinning regime we thus observe an effect reminiscent of active thinning~\cite{burkholder20, jayaram23}.

Having access to the force allows for further insight into the energetics of a moving, deforming body. Ignoring the dissipation incurred by the APs~\cite{bebon25} and solvent~\cite{nasouri21}, and focusing on the energetics of the flagellum, the two relevant contributions are the rate of work $\dot W_\text{def}$ (Eq.~\eqref{eq:W_def}) to deform the body and the rate $\dot W_\text{drag}$ (Eq.~\eqref{eq:W_drag}), either applied to counteract the drag ($>0$) or recuperated through interactions with APs ($<0$) as the body is moved at constant speed $u$ (see End Matter for details).

In case the body swims freely with speed $u_\ast$, the drag force vanishes and the average total work rate becomes $\langle \dot W \rangle = \IInt{t}{0}{\tau} \dot W_\text{def}/\tau$. Since it depends only on the flagellum's material properties and the oscillation frequency $\om$ (Eq.~\eqref{eq:W_def}), the thermodynamic cost to drive the periodic deformation is independent of the degree of bath activity $v_0$. Accordingly, a freely swimming flagellum is neither assisted nor hindered by the presence of active bath constituents and occurs at constant speed $u_\ast$ under injection of work $\langle \dot W \rangle$. Since $u_\ast \propto \om$ (Fig.~S4(a) in \cite{sm}), tuning the swim speed $u_\ast$ requires a change in the beating frequency, which is reflected in the work rate $\langle \dot W \rangle \propto \om^3$ (obtained by integrating Eq.~\eqref{eq:W_def}). Interestingly, the different scaling behaviors suggest that the work-to-speed conversion is most effectively performed in the small-$\om$ regime, where the linear growth of $u_\ast$ exceeds the cubic scaling of $\langle \dot W \rangle$.

The situation changes if the flagellum is subjected to an external load, e.g., a constant force that opposes its motion. At steady state, forces balance and the body extracts work from the active bath $\langle \dot W_\text{drag} \rangle <0$ to counteract the applied force; $\langle \dot W_\text{load} \rangle = -\langle \dot W_\text{drag} \rangle$. Unlike in the free case, however, the body's speed is now sensitive to bath properties since $v_0$ determines the amplitude of drag forces (see Fig.~\ref{fig:force_sim}(c)). Higher swim speeds $u_\ast$ under external load are achievable by tuning the self-propulsion speed of APs, while the energy required to deform the body remains the same.

In summary, we have experimentally investigated the friction forces on a flagellum-shaped body undergoing both beating and translation in a suspension of light-activated active particles (APs). When the translation speed is below the flagellum’s free-swimming speed, it experiences a negative drag force which supports its translational motion. At higher speeds, the drag becomes positive and asymptotically approaches a Stokes-like regime with a constant, speed-independent friction coefficient. This behavior arises from a spatiotemporal density distribution of APs which preferentially accumulate at locations facilitating propulsion.
The measured forces between the virtual flagellum-shaped object and the APs are of a similar order of magnitude as the viscous drag that such an object would experience, see SM \cite{sm} for details.
Our findings are supported by numerical simulations and an analytical model, which can be extended to predict mean forces on slender bodies in active fluids at both low and high transverse velocities.


\textit{Acknowledgments.}---CB acknowledges funding by the ERC AdG. BRONEB (Grant No. 101141477) and the Centre of Excellence 2117  (Grant No. 422037984). RB and TS acknowledge funding by the DFG through the collaborative research center TRR 146 (Grant No. 404840447). The authors declare no competing interests.


\textit{Data availability.}---The data that support the findings of this Letter are openly available ~\cite{data}.


\bibliography{references_full}


\onecolumngrid
\section*{End Matter}
\twocolumngrid
\renewcommand{\theequation}{A\arabic{equation}}
\setcounter{equation}{0} 

\emph{Measurement of forces.}---To determine values for $\langle \vec{F} \rangle$, as defined in Eq.~\eqref{eq:force:exp}, we measure $\vec{F}_{\text{rep},i}(t)$ for all interacting APs $i$ at a fixed time $t$, and average over an integer multiple of the object's deformation period. In the experiments, we have direct access to $\vec{F}_{\text{rep},i}(t)$ since both the position and shape of the virtual object, as well as the APs' positions, are controlled and recorded in real time. In other words, we know the position, orientation, and strength of all applied light gradients that constitute the repulsive interaction between APs and the light-induced object---each gradient applied to the $i$th AP contributes to $\vec{F}_{\text{rep},i}(t)$. Following the relation $\vec{F}_\text{rep}=-f_0 \left( \vec{\nabla}I/|\vec{\nabla}I| \right)$ given in the main text, it is apparent that $\vec{F}_{\text{rep},i}(t)$ aligns with the direction of the applied light gradient. Experimentally, this gradient is generated by a row of five laser spots placed near each corresponding AP and oriented parallel to the surface of the interacting object. The magnitude, $f_0 = 650 \kT/\sigma$, has been determined in separate experiments, as described in the Supplemental Material \cite{sm}. For averaging, $\vec{F}_\text{{rep},i}(t)$ is normalized by the instantaneous AP density to facilitate comparison across measurements. For further details on the normalization, averaging and determination of $\langle \vec{F} \rangle$ we refer to the Supplemental Material \cite{sm}.

\emph{Flagellum energetics.}---The elastic deformation of the flagellum is externally driven by applying mechanical work on the body. Following the time-periodic protocol $y(x,t)$ (Eq.~\eqref{eq:flagellum_model}), deformation work is performed at a rate of
\begin{align}
    \dot W_\text{def} = \IInt{x}{0}{L} \fd{U_\text{def}}{y} \pd{y}{t} 
    \label{eq:W_def}
\end{align}
and stored in the elastic energy of the deformation. Using classical elastic beam theory~\cite{landau09}, the internal energy associated with a specific flagellum configuration can be modeled through
\begin{align}
  U_\text{def} = \frac{1}{2} EI \IInt{x}{0}{L} |\kap|^2 \approx -\frac{1}{2} EI \IInt{x}{0}{L} |y''|^2,
\end{align}
absorbing material properties into the body's Young's modulus $E$ and its second moment of area $I$. To additionally move the flagellum at a constant speed $u$ against the drag force density $\vec f(x,t)$ requires mechanical work at the rate 
\begin{align}
    \dot W_\text{drag} = \IInt{x}{0}{L} \left(\pd{y}{t} \vec e_y - u \vec e_x \right)  \cdot \vec f
    \label{eq:W_drag}
\end{align}
that is either applied to ($\dot W_\text{drag} > 0$), or extracted from ($\dot W_\text{drag} < 0$) the system. Note that another contribution to the overall dissipation stems from the active bath since APs require a steady influx of energy to fuel their non-equilibrium dynamics~\cite{bebon25}. We neglect this contribution here and focus on the deformable body.

\clearpage
\newpage
\renewcommand{\thesection}{S\arabic{section}}
\renewcommand{\thefigure}{S\arabic{figure}}
\renewcommand{\theequation}{S\arabic{equation}}
\setcounter{equation}{0}
\setcounter{figure}{0}
\setcounter{page}{1}
\setcounter{section}{0}

\onecolumngrid
\begin{center}\textbf{\large{\small Supplemental Material to} \\ Negative drag force on beating flagellar-shaped bodies in active fluids}\\[0.5cm]
Timo Knippenberg$^1$, Robin Bebon$^2$, Thomas Speck$^2$, and Clemens Bechinger$^{1,3}$\\[0.2cm]
\emph{$^1$Fachbereich Physik, Universität Konstanz, 78464 Konstanz, Germany}\\
\emph{$^2$Institute for Theoretical Physics IV, University of Stuttgart, Heisenbergstr.~3, 70569 Stuttgart, Germany}\\
\emph{$^3$Centre for the Advanced Study of Collective Behaviour, Universtätsstraße 10, 78464 Konstanz, Germany}
\\[1cm]\end{center}

\twocolumngrid

\subsection{ Experimental setup}

\subsubsection{Self-propulsion mechanism}

We perform the experiments with active Janus colloids, specifically SiO$_2$-spheres with $\sigma = \SI{7.89}{\micro\meter}$ diameter which are coated with a layer of $\SI{60}{\nano\meter}$ carbon on one hemisphere. The particles are suspended in a binary mixture of water and propylene glycol n-propyl ether (PnP) (40\% m) and sedimented to the bottom of a sample cell (\textit{Hellma}, $\SI{200}{\micro\meter}$ height), which is kept at $\SI{28}{\celsius}$, rendering the fluid sub-critical.
When illuminated with laser light, the carbon caps are selectively heated which is causing local demixing of the surrounding critical fluid. As a result, the particles self-propel with the uncapped hemisphere in front \cite{RN5,RN2}.
To achieve individual control of each particle, each AP's carbon cap (center of laser spot is displaced from the APs' centers of mass by $\SI{1.25}{\micro\meter}$ into the direction of the carbon cap) is illuminated by a green ($\lambda = \SI{532}{\nano\meter}$) laser spot with approximately $\SI{4.5}{\micro\meter}$ beam waist, by rapidly scanning a laser beam with a power of up to $\SI{6}{\milli\watt}$ over all AP positions. The rapid scanning over all AP positions is realized using an acoustic-optical deflector which is connected to a computer, performing real-time particle detection through a connected microscope.
In total, this forms a feedback-control where the illuminating laser spot is following the APs' trajectories. Since the laser scanning occurs on a much faster timescale than that of the fluid demixing dynamics, the illumination appears quasi-stationary. As a result, the APs perform a 2D active Brownian particle-like motion with a persistence time of $\tau_\text{p} \approx \SI{800}{\second}$ and a variable speed \cite{RN7}, the latter depending on the intensity of the laser spot. In our experiments, we realize AP velocities up to $v_0 \approx \SI{0.2}{\micro\meter\per\second}$. Hydrodynamic interactions between the APs and the substrate constrain the poles of the APs such that they are parallel aligned to the substrate. This confines the effective rotational diffusion to 2D \cite{RN9}.

\subsubsection{Realization of flagellum-shaped objects}

To evoke the interaction between the virtual flagellum and the APs, we make use of the fact that this kind of active colloids display negative phototaxis when placed in a light gradient \cite{RN3,RN4}. Utilizing our feedback-controlled setup, we place additional laser spots at the edge of every AP that has occupied a position in the field-of-view which is considered to be \textit{inside} the flagellum. These additional light spots create a local light gradient close to the corresponding APs, which propels them with velocity $v_\text{rep}$. The light gradient is chosen to be oriented normal to the flagellum contour and thus the induced active velocity is directed away from the flagellum. Technically, the gradient is generated by superimposition of 5 overlapping ($\SI{1.25}{\micro\meter}$ distance between spot centers) laser spots, each of power $I$. These 5 spots form a line which follow the contour of the flagellum. The closest distance between this line and the APs center of mass is held constant at $\SI{2.75}{\micro\meter}$. 

\begin{figure}[t]
    \centering
    \includegraphics{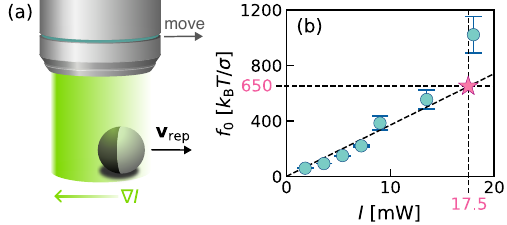}
    \caption{(a) Sketch of the experimental setup to measure the force magnitude, by exploiting the negative phototaxis of APs within local light gradients. (b) Measured values for the force magnitude of $\vec{F}_\text{rep}$ in dependence on the peak intensity $I$ of the driving light gradient. The lines are guides to the eye. The star indicates the values we used throughout the study.}
    \label{fig:SM_experiment}
\end{figure}

As shown by previous studies \cite{RN7,RN8}, this implementation can be mapped to the presence of a linear repulsive potential, effectively pushing away APs from the boundary of the flagellum. 
For example, it has been experimentally shown that APs being exposed to such a light-induced constant force exhibit a exponentially decaying sedimentation profile~\cite{RN7}. This is in line with theoretical predictions of ABPs in a constant force field~\cite{RN63}. We deduce that our method of barrier-creation is based on the imposition of physical forces on the APs, although they are light-induced.

We calibrate our method of boundary-creation by identifying the magnitude $f_0$ of the repulsive force $\vec{F}_\text{rep}$ induced by the light gradient. For this purpose, we have performed supplementary experiments, in which the Janus particles were exclusively subject to a continuously applied (local) light gradient, giving rise to a motion along a straight line (see Fig.~\ref{fig:SM_experiment}(a)). That means that there was no additional laser spot to drive the particles in an ABP-like fashion in these specific supplementary experiments. By measuring the particle velocity $v_\text{rep}$, one can estimate the repulsive force $\vec{F}_\text{rep}$ \emph{via} the particles' mobility as
\begin{equation}
    \vec{F_\text{rep}} = 6 \pi \eta \sigma \vec{v}_\text{rep},
\end{equation}
where $\eta = \SI{4}{\milli\pascal\second}$ is the viscosity of the water-PNP-mixture.
Figure~\ref{fig:SM_experiment}(b) shows the measured magnitude $f_0$ of $ \vec{F_\text{rep}}$ in dependence on the peak laser power $I$ in the light gradient. For this study, we have chosen the maximum power $I$ of the light gradient to be $\SI{17.5}{\milli\watt}$ in all our experiments, which corresponds to a repulsive force of roughly $f_0 \approx 650~\kT/\sigma$.

\begin{figure}[t]
    \centering
    \includegraphics{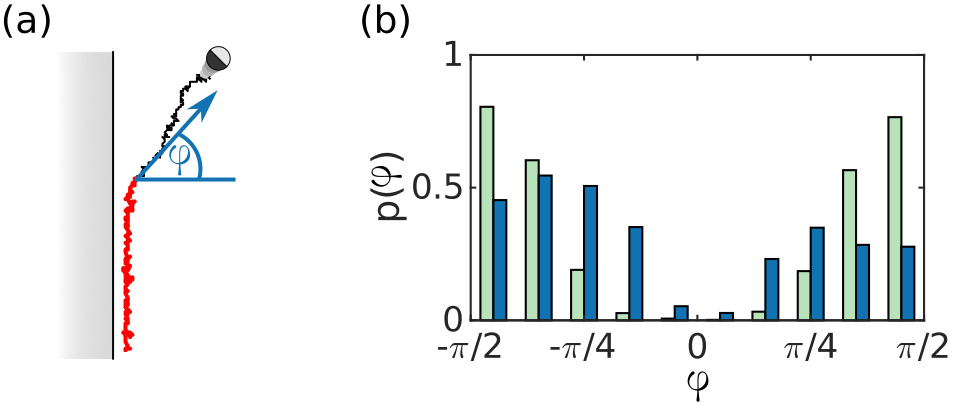}
    \caption{(a) Illustration of the departure angle $\vhi$ for an AP moving away from a flat wall, the latter created by a light gradient. During the initial part of the trajectory (red) where the AP moves along the wall, the AP is considered to be at the boundary (b) Probability distribution of AP departure angles normal to the confining boundary. The blue bars are experimentally measured data, the green bars are results from corresponding Brownian dynamics simulations.}
    \label{fig:SM_scattering}
\end{figure}

To affirm that the light gradient primarily influences the APs' position and not their orientation, we perform additional supplementary experiments, in which we measure the departure angles $\vhi$ in which APs move away from a hard wall implemented using light gradients as described above, i.e. in the same way as we create the virtual flagellum-shaped object. For this, we closely follow a procedure described in \cite{RN44}, which was originally developed to determine the scattering angles of phototactic bacteria at a light-induced boundary. Specifically, we confined a dilute suspension of APs with $v_0 \approx \SI{0.19}{\micro\meter}$ into a rectangular shaped box of size $\SI{200}{\micro\meter} \times \SI{120}{\micro\meter}$, which is smaller than their typical persistence length. We record the angles $\vhi$ in which APs depart from the confining boundaries, measured relative to the boundaries normal vector pointing inwards (cf. Fig.~\ref{fig:SM_scattering}(a)). We count APs as \textit{departed} when they are displaced at least $\SI{1}{\micro\meter}$ from the closest boundary. Figure~\ref{fig:SM_scattering}(b) shows the experimentally measured probability distribution of $\vhi$ (dark), together with results from Brownian dynamics simulations (bright) of the same system, where the confining boundaries are modeled as an external force with constant magnitude. For the experiments as well as for the simulations, APs mostly depart from the boundary in a shallow angle. These results confirm that wall departure is caused by rotational diffusion and reorientational torques, which would induce departure angles centered around $\vhi = 0^\circ$, are absent.

Supplementary Video 1 is the microscopy recording of an exemplary measurement on which an animation of the flagellar-shaped object is superimposed. The measurement parameters are $v_0 = \SI{0.1}{\micro\meter\per\second}$, $\omega = \SI{0.013}{\hertz}$, $u = \SI{0.2}{\micro\meter\per\second}$. The video is accelerated by a factor of 300. The field of view has a size of $\SI{600}{\micro\meter} \times \SI{500}{\micro\meter}$

\subsubsection*{Measurement parameters}

Each experimental data point shown is the result of several measurements, in total this adds up to at least $\SI{5.5}{\hour}$ of experiments per data point. As the accessible range for transversal motion of the sample cell is limited, more but shorter experiments are performed for large $u$, compared to less and longer experiments for small $u$. To give the system time to relax into a non-equilibrium steady state, the first $\SI{15}{\minute}$ of trajectories in each measurement are discarded. Microscope images are recorded with a frame rate of $\SI{2}{\per\second}$. Due to thermal noise in our optical setup, the positioning of laser spots has to be carefully calibrated before every measurement.

The area packing fraction $\phi$ of APs varies both between experiments and within individual experimental runs due to the limited field of view. The average is $\phi = 0.03$ with deviations up to $\pm 25~\%$ in extreme cases. These density fluctuations complicate direct comparisons of $\langle F_x \rangle$ across different experiments and simulations. To mitigate this issue, we normalize $\langle F_x \rangle$ by the instantaneous packing fraction relative to the average density. Specifically, for each measurement and simulation, we determine the normalized force $\langle F_x \rangle = (\phi(t)/\phi)\left\langle F_{x,\text{meas}}\right\rangle$, where $F_{x,\text{meas}}$ is the directly measured force and $\phi(t)$ is the area packing fraction at time $t$. We verified that within this regime—well below the onset of motility-induced phase separation—the dependence of $\langle F_{x,\text{meas}} \rangle$ on $\phi$ is linear, thereby allowing this normalization approach.

\subsubsection*{Details on averaging}

\begin{figure}[t]
    \centering
    \includegraphics{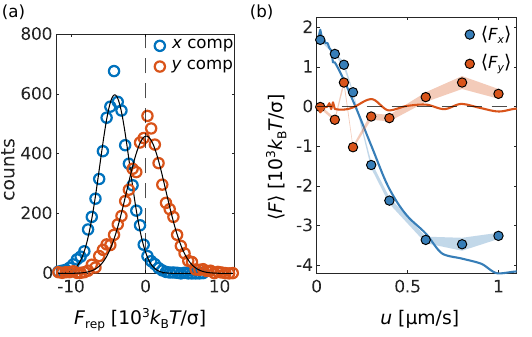}
    \caption{(a) Exemplary measurement with the parameters $\omega = \SI{0.0016}{\per\second}$, $u = \SI{0.3}{\micro\meter\per\second}$ and $v_0 = \SI{0.19}{\micro\meter\per\second}$, showing the distribution of force components $[\vec{F}_{\mathrm{rep},i}]_x$ and $[\vec{F}_{\mathrm{rep},i}]_y$ measured over multiple beating periods. The black lines are fits to a Gaussian function. (b) $x$- and $y$-component of the average force that APs exert on the flagellar-shaped body (cf.~Fig.~2) for $\omega = \SI{0.013}{\micro\meter\per\second}$ and $v_0 = \SI{0.19}{\micro\meter\per\second}$. Circles display experimental data, with standard deviation of the mean indicated by the shaded areas. Solid lines represent the corresponding force components determined from simulations.}
    \label{fig:y-forces}
\end{figure}

In experiments, we record the instantaneous forces exerted by individual APs over multiple periods of the body's beating. For a representative experiment, with $\omega = \SI{0.0016}{\per\second}$, $u = \SI{0.3}{\micro\meter\per\second}$ and $v_0 = \SI{0.19}{\micro\meter\per\second}$, we depict the resulting force distributions (i.e. one for the $x$- and one for the $y$-component of $-\vec{F_\text{rep}}$) in Fig.~\ref{fig:y-forces}(a). The average force components, $\langle F_x \rangle$ and $\langle F_y \rangle$, are obtain as the arithmetic means. To estimate the statistical uncertainty, we fit each force distribution with a Gaussian (black lines in Fig.~\ref{fig:y-forces}(a)) and extract its standard deviation.

While the dependence of $\langle F_x \rangle$ on the experimental parameters is extensively covered in the main text, we find that $\langle F_y \rangle \approx 0$ for all parameters, as expected for the symmetric deformation along the $y$-direction. Notably, any deviations from zero are small and appear to be caused by fluctuations that follow no visible trend. A representative data set is shown in Fig.~\ref{fig:y-forces}(b) for $\omega = \SI{0.013}{\per\second}$ and $v_0 = \SI{0.19}{\micro\meter\per\second}$. The $y$-component of the force fluctuates around zero for both the experimental (circles) and simulation (lines) data.

\subsection{Simulation details and theoretical considerations}

\subsubsection*{Simulation setup}

We perform Brownian dynamics simulations in two dimensions. The position $\vec{r}_i$ and orientation $\vec{\vartheta}_i$ of the $i$-th AP are governed by
\begin{align}
    \vec{r}_i(t+\Delta t) &= \vec{r}_i(t) + v_0 \vec{e}_i \Delta t +\\
    \frac{\Dt}{\kT} &\left(-\vec{\nabla} V_\text{WCA} + \vec{F}_{\text{rep},i}  \right) \Delta t + \sqrt{2 \Dt  \Delta t} \vec{\xi}_{\text{t},i} \notag \\
    \vartheta_i(t + \Delta t) &= \vartheta_i(t) + \sqrt{2 \Dr \Delta t} \xi_\text{r},
\end{align}
where $\vec{e} = ( \cos(\vartheta), \sin(\vartheta) )^\top$ and $\vec{\xi}_{\text{t},i}$ and $\xi_\text{r}$ are stochastic variables with zero mean and unit variance. The duration of an integration step is $\Delta t = \SI{0.1}{\second}$. The APs interact with each other via the repulsive Weeks-Chandler-Anderson-potential $V_\text{WCA}$. The interaction with the flagellum is included via $\vec{F}_\text{rep}$, as described in the main text. However, we have chosen a larger value of $f_0 = 2000 \kT/\sigma$, whereas in the experiments $f_0 = 650 \kT/\sigma$. This gives access to larger values of $\vec{u}$ in the simulations, because for $u \gtrsim \SI{1}{\micro\meter\per\second}$ a force magnitude of $f_0 = 650 \kT/\sigma$ is not strong enough to keep APs outside of the flagellum. We have carefully checked that the influence of $f_0$ on $\langle F_x \rangle$ is negligible as long as its large enough to keep APs outside the flagellum.

Simulation parameters have been chosen to fit to the experiments, namely $\Dt = \SI{1.4e-14}{\meter\squared\per\second}$, $\Dr^{-1} = \SI{800}{\second}$, and a packing fraction of $\phi=0.03$. The simulation box size is $\SI{1250}{\micro\meter} \times \SI{600}{\micro\meter}$. The transversal velocity of the flagellum has been included by adding a constant drift velocity $-u$ to the APs. Consequently, the flagellum is kept at a constant position inside the simulation box with its front end at $x = \SI{663}{\micro \meter}$. We have chosen periodic boundary conditions in $\pm y$-direction. In $x$-direction, periodic boundary conditions could lead to problems, since the AP density modulations generated by the flagellum might propagate \textit{through} the periodic boundary and thus reappear in front of the flagellum. To circumvent this problem, we have instead chosen to constantly create new APs in the region $x > \SI{1200}{\micro\meter}$ until the target packing fraction is reached, and remove them from the simulation box when their $x$-coordinate is negative. Each simulation data point represents a single simulation run with 400,000 time steps. Supplementary Video 2 provides an animation of an exemplary simulation run matching the parameters for the experiment in Supplementary Video 1.

\subsubsection{Numerical validation of theoretical ansatz}

\begin{figure}[t]
    \centering
    \includegraphics{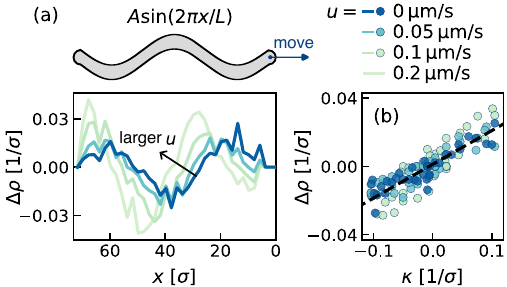}
    \caption{(a) Simplified probe shape and density difference profiles for different pulling speeds. (b) Density difference plotted over the curvature for different speeds $u$. The dashed line depicts a joint linear fit of all data points with slope $\zeta\approx 0.2$. We used $\omega = 0$ and $v_0 = \SI{0.19}{\micro\meter\per\second}$ throughout.}
    \label{fig:density_diff}
\end{figure}

To validate the main assumption of the minimal model, we perform additional simulations and obtain the density difference along a simple sinusoidal probe (Fig.~\ref{fig:density_diff}(a)), with amplitude $A = 5 \sig$, (wave-)length $L = 63 \sig$, and width $d = 4.5 \sig$. In the stationary case ($u=0$) we verify to good agreement the linear relation between density difference and curvature, even for the repulsive bath particles considered here (Fig.~\ref{fig:density_diff}(b)). 

Upon pulling the probe along the $-x$-direction, density profiles visibly shift in the opposing direction (Fig.~\ref{fig:density_diff}(a)), the extent of which increases linearly with the pull speed $u$. As illustrated in Fig.~4(a) of the main text, this shift can be rationalized by considering the influence of local curvature changes on the, comparably slow reorienting, active bath particles. Accounting for the shift in position $x'=x+u\delta t$ once the flagellum is driven, we further validate the linear relation $\Delta \rho \approx \zeta \kap(x',t')$ in Fig.~\ref{fig:density_diff}(b). Note that the slope $\zeta$ is (almost) identical to the passive case.

\subsubsection{Swim speed and drag coefficient}

\begin{figure}[!b]
  \centering
  \includegraphics[scale=1]{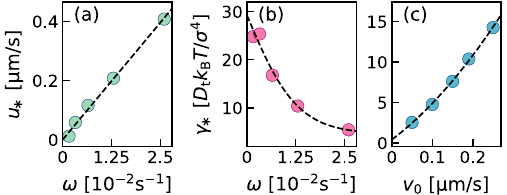}
  \caption{(a) Swim speed $u_\ast$ as a function of deformation frequency. The black dashed line depicts a linear fit. Drag coefficient $\gam_\ast$ (Eq.~\eqref{eq:drag_stall}) as a function of (b)~frequency $\om$ and (c)~bath self-propulsion speed $v_0$. Dashed lines are a guide to the eye.}
  \label{fig:susc_stall}
\end{figure}

The predicted drag force $\langle F_x\rangle=F_\text{def}-\gam_\ast u$ (Eq.~(4)) on the flagellum can be decomposed into a contribution
\begin{equation}
    F_\text{def} = \frac{f_0 \zeta}{\tau} \IInt{t}{0}{\tau} \IInt{x}{0}{L} \left[ \kap - \delta t (\partial_t \kap +c\partial_x \kap) \right] \sin\theta
\end{equation}
arising due to the deformations and the drag coefficient
\begin{equation}
    \gam_\ast \equiv - \frac{\zeta f_0 \delta t}{\tau} \IInt{t}{0}{\tau} \IInt{x}{0}{L}  (\partial_x \kap)\sin \theta
    \label{eq:drag_stall}
\end{equation}
close to the swim speed $u_\ast=F_\text{def}/\gam_\ast$ corresponding to the force balance $\langle F_x\rangle=0$. Its values, obtained from the fits in Fig.~4(b,c) of the main text, are plotted in Fig.~\ref{fig:susc_stall}(b,c) as a function of beating frequency $\om$ and AP speed $v_0$. The drag coefficient reduces monotonically with $\om$ and approaches a non-zero plateau value (Fig.~\ref{fig:susc_stall}(b)) for high beating frequencies. Any particles trying to enter the envelope of the oscillation are immediately ejected, whence on the timescale of AP diffusion the object appears rigid. Conversely, the drag coefficient steadily increases as a function of $v_0$ and follows a weakly non-linear trend that is best fitted by a second-order polynomial (Fig.~\ref{fig:susc_stall}(c)). We thus observe ``active thickening'' close to the swim speed ($u\simeq u_\ast$) and active thinning in the shear-thinning regime ($u\to\infty$).

\subsubsection{Influence of the flagellum's geometry}

\begin{figure}[t]
    \centering
    \includegraphics{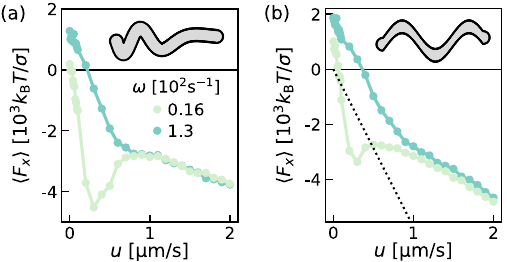}
    \caption{Speed-force curves for different periodically undulating shapes at various $u$ and $\omega$, and fixed $v_0 = \SI{0.19}{\micro\meter\per\second}$. (a) Flagellum according to Eq.~(1), but with constant tangential thickness. (b) Sinusoidal shape with constant amplitude. Shapes are indicated in the respective insets. The dotted black line indicates the drag experienced by a similarly sized sphere that is dragged through a binary water-PNP mixture.}
    \label{fig:differentShapes}
\end{figure}

To demonstrate that our results are not constraint to the specific choice of the flagellum's shape, we have repeated the simulations with slightly altered shapes.
Specifically, we calculated $\langle F_x \rangle$ for a pure sinusoidal wave with constant amplitude and for the model as in Eq.~(1), but with constant thickness normal to the edge, rather than in $y$-direction as previously. Figure~\ref{fig:differentShapes} shows $\langle F_x \rangle$ together with the corresponding shapes. As apparent, the qualitative features of $\langle F_x \rangle$ are robust to slight changes in the flagellum's geometry, provided some key features are retained: (i) A periodic undulation that propagates towards the back of the flagellum is required to observe a sign change in $\langle F_x \rangle$, and (ii) concave regions along the body's contour are mandatory for the formation of local minima in $\langle F_x \rangle$.

The dimensions of the virtual flagellum used in the main text were chosen to resemble the order of magnitude of relative lengths of an eukaryotic flagellum described in \cite{rikmenspoel65}, where $A_0/L \approx 0.08$ and the maximum amplitude is approximately doubled after each full wavelength, but enlarged by roughly a factor of 4.5. This is to better sample the size and persistence length of our APs.\\

\subsubsection{Rough estimation of the drag force magnitude}

To obtain a rough estimate of the order of magnitude of the measured effects, we have plotted in Fig.~\ref{fig:differentShapes}(b) the estimated viscous drag force of a similar sized (radius $4 A_0 = \SI{40}{\micro\meter}$) sphere moving trough the solvent of water-PNP (viscosity $\eta = \SI{4}{\milli\pascal\second}$) using Stokes' law.
As apparent, both effects are of similar magnitude in case of fixed AP area fraction of $\phi \approx 0.03$. Note however, that the magnitude of the drag force induced by the APs is expected to scale linearly with $\phi$ up to a certain threshold above which many-particle effects (e.g. MIPS) produce a more complex behavior.

\subsection*{Supplementary movies}

\textbf{SupplementaryMovie1:}
Microscope recording of a representative measurement. The flagellar shaped object is superimposed as an animation. The measurement parameters are $v_0 = \SI{0.1}{\micro\meter\per\second}$, $\omega = \SI{0.013}{\hertz}$ and $u = \SI{0.2}{\micro\meter\per\second}$. The video is accelerated by a factor of 300. The field of view has a size of $\SI{600}{\micro\meter} \times \SI{500}{\micro\meter}$.

\textbf{SupplementaryMovie2:}
Animation of a representative simulation run. The simulation parameters are  $v_0 = \SI{0.1}{\micro\meter\per\second}$, $\omega = \SI{0.013}{\hertz}$ and $u = \SI{0.2}{\micro\meter\per\second}$, i.e.~they match the parameters of the experiments in \textit{SupplementaryMovie1}. The animation is rendered with a frame rate of $\SI{3}{\per\second}$, but only every 500th integration step of the simulation is plotted.
The $x$-axis and the $y$-axis are scaled in units of $\sigma$. The APs marked as red are inside the area in which new APs are constantly created.

\end{document}